\documentclass[12pt, preprint]{aastex}
\usepackage{graphics}

\shorttitle{ISO observations of the \object{Hickson 31} group}
\shortauthors{O'Halloran et~al.}

\begin{document}

\title{ISO observations  of Hickson Compact Group 31  with the central
Wolf-Rayet galaxy NGC 1741}

\author{B.  O'Halloran\altaffilmark{1}, L. Metcalfe\altaffilmark{1,2},
B.      McBreen\altaffilmark{1},      R.     Laureijs\altaffilmark{2},
K.      Leech\altaffilmark{2},      M.      Delaney\altaffilmark{1,3},
D. Watson\altaffilmark{1,4} and L. Hanlon\altaffilmark{1}}

\email{halloran@bermuda.ucd.ie} 
\email{lmetcalf@iso.vilspa.esa.es}
\email{brian.mcbreen@ucd.ie}
\email{rlaureijs@esa.int}
\email{kleech@iso.vilspa.esa.es}
\email{mdelaney@bermuda.ucd.ie}
\email{dwatson@bermuda.ucd.ie}
\email{lhanlon@bermuda.ucd.ie}

\altaffiltext{1}{Physics Department, University College Dublin, Dublin
4,   Ireland}   \altaffiltext{2}{Research   and   Scientific   Support
Department  of  ESA, Villafranca,  P.O.   Box  50727, E-28080  Madrid,
Spain}     \altaffiltext{3}{Stockholm    Observatory,     SE-133    36
Saltsj\"{o}baden,  Sweden}   \altaffiltext{4}{X-Ray  Astronomy  Group,
Dept. of Physics and Astronomy, University of Leicester, Leicester LE1
7RH, United Kingdom}

\begin{abstract}

Hickson Compact  Group (HCG) 31,  consisting of the  Wolf-Rayet galaxy
NGC 1741  and its irregular  dwarf companions, was observed  using the
Infrared  Space  Observatory.   The  deconvolved ISOCAM  maps  of  the
galaxies using  the 7.7 $\mu$m and  14.3 $\mu$m (LW6  and LW3) filters
are  presented,  along  with   ISOPHOT  spectrometry  of  the  central
starburst  region of  NGC  1741 and  the  nucleus of  galaxy HCG  31A.
Strong  mid-IR emission  was detected  from the  central burst  in NGC
1741, along with strong PAH features and a blend of features including
[S IV]  at 10.5  $\mu$m.  The 14.3/6.75  $\mu$m flux ratio,  where the
6.75 $\mu$m flux was synthesized from the PHT-S spectrum, and 14.3/7.7
$\mu$m flux ratios suggest that  the central burst within NGC 1741 may
be  moving   towards  the  post-starburst   phase.   Diagnostic  tools
including the ratio of the integrated  PAH luminosity to the 40 to 120
$\mu$m infrared  luminosity and  the far-infrared colours  reveal that
despite the high surface brightness  of the nucleus, the properties of
NGC 1741 can  be explained in terms of a starburst  and do not require
the presence  of an  AGN.  The Tycho  catalogue star  TYC 04758-466-1,
with m$_{V}$ =  11.3 and spectral type F6, was  detected at 7.7 $\mu$m
and 14.3 $\mu$m.
\end{abstract}
\clearpage

\section{Introduction}

Compact  groups  are  small  systems  of  galaxies  with  each  member
apparently in close proximity to the other members.  They are presumed
to be physically associated and often show morphological peculiarities
indicative of gravitational interaction.  Hickson (1982), by examining
Palomar Observatory  Sky Survey plates, identified  100 compact groups
of  galaxies with membership  defined by  the following  criteria: (i)
There are more than 4 galaxies in the group within 3 magnitudes of the
brightest member.  (ii) The group must be isolated so that no external
galaxies fall  within a radius  of 3 times  the angular extent  of the
group.  (iii) The  total magnitude per arcsec\(^{2}\) of  the group is
less  than 26.   Considering  their high  densities  and low  velocity
distributions, compact  groups provide an opportunity  to study galaxy
interactions  and their  effects, especially  considering  their short
dynamical lifetimes of $\sim$ 10\(^{9}\) yr \citep{rub90}.

One  facet  of such  interactions  is  the  presence of  intense  star
formation,  and  in particular  the  presence  of Wolf-Rayet  features
within the spectra of these objects.  Wolf-Rayet galaxies are a subset
of  emission line  and HII  galaxies whose  integrated spectra  have a
broad  emission feature  around 4650\AA~which  has been  attributed to
Wolf-Rayet stars.  The emission feature usually consists of a blend of
lines,  namely  HeI  $\lambda$4686,  CIII/CIV  $\lambda$4650  and  NIII
$\lambda$4640.  The  CIV $\lambda$5808 line  can also be  an important
signature of  Wolf-Rayet activity.  Wolf-Rayet  galaxies are important
in  understanding  massive  star  formation  and  starburst  evolution
\citep{sch98}.  The  Wolf-Rayet phase in massive stars  is short lived
and hence  gives the possibility  of studying an  approximately coeval
sample  of  starburst  galaxies.   The  first  catalog  of  Wolf-Rayet
galaxies was  compiled by  Conti (1991) and  contains 37  galaxies.  A
large number  of additional  Wolf-Rayet galaxies have  been identified
and  a recent  catalog containing  139 galaxies  has been  compiled by
Schaerer et~al.   (1999).  Wolf-Rayet galaxies are found  among a wide
variety of morphological types, from blue compact dwarfs \citep{met96}
and irregular  galaxies, to massive spirals and  luminous merging IRAS
galaxies  \citep{rig99},  with  a  range  of  metallicities  and  ages
\citep{cer94,mas99}.

   Hickson  Compact Group  (HCG) 31  consists  of 8  galaxies and  was
observed  using  the   Infrared  Space  Observatory\footnote{Based  on
observations with ISO,  an ESA project with instruments  funded by ESA
Member  States  (especially the  PI  countries:  France, Germany,  the
Netherlands and the United Kingdom) and with the participation of ISAS
and NASA.}  (Fig.  1).  Hickson  (1982) identified galaxies A to D from
the Palomar Sky  Survey in order of luminosity, with  A being the most
luminous because it included the central starburst.  However follow-up
surveys  \citep{rub90,con96,igl97,joh99,joh00} included  the starburst
in galaxy C rather than A  thus making C the most luminous member, and
we  follow this  convention.  Galaxies  A  and C  are also  designated
together as NGC 1741, one  of the most luminous Wolf-Rayet galaxies in
the catalog of Conti (1991).   Rubin et al.  (1990) identified several
further members of  the HCG 31 group, galaxies E, F,  G and a possible
member further  north, Q.  The galaxies  C and G were  included in the
Markarian  survey  and  were  designated  as Mrk  1089  and  Mrk  1090
respectively.  HCG 31 is situated  at a distance of 54.8 Mpc, assuming
a Hubble  parameter of 75 km sec\(^{-1}\)  Mpc\(^{-1}\).  The redshift
of  galaxy  D  \citep{hic92} places  it  at  a  distance of  359  Mpc,
indicating  it is not  a bona  fide member  of the  HCG 31  group.  We
present ISOCAM and  ISOPHOT results on HCG 31 as  part of an observing
programme   addressing    several   galaxies   exhibiting   Wolf-Rayet
signatures.   The observations  and  data reduction  are presented  in
Sect.   2.  The results  are contained  in Sect.   3 and  discussed in
Sect.  4.  The conclusions are summarised in Sect.  5.

\section{Observations and Data Reduction}

The   ISO   \citep{kes96}  observations   were   obtained  using   the
mid-infrared camera ISOCAM \citep{ces96} and the spectrometric mode of
the  ISO  photopolarimeter  ISOPHOT \citep{lem96}.   The  astronomical
observing  template  (AOT)  used  was CAM01  \citep{blom01},  for  the
imaging observations  and PHT40, for spectrometry.   The observing log
of the CAM01 observations at 7.7  $\mu$m and 14.3 $\mu$m using the LW6
and LW3 filters and the PHT40 observations is presented in Table 1.

\subsection{ISOCAM}

Each  ISOCAM observation  had the  following configuration:  3$''$ per
pixel field of view (PFOV), integration  time of 5.04 s per readout, a
raster with $4\times4$ spacecraft  pointings with a different stepsize
on  each axis:  24$''$  stepsize  between raster  points  and a  6$''$
stepsize  between  raster lines.   The  ISOCAM  detector  was used  in
association with  one of  a selection of  four imaging  lenses.  These
allowed plate-scales of 1.5$''$, 3$''$,  6$''$ and 12$''$ per pixel to
be  chosen, resulting in  turn in  overall detector  fields-of-view of
48$''$,   96$''$,  and   192$''$   square.   The   12$''$  per   pixel
field-of-view (12$''$ PFOV) did not access much more sky area than the
6$''$ PFOV due to vignetting  of the instrument beam outside a nominal
maximum  unvignetted  field having  3$'$  diameter.  The  observations
discussed here  employed the 3$''$ per pixel  field-of-view, giving an
overall detector  field of  view of 96$''$  square.  Apart  from these
details of  the detector plate-scale  and FOV, the  spatial resolution
was  determined  by the  point  spread  function  (PSF), which  had  a
full-width in arcseconds at  the first Airy minimum of 0.84x$\lambda$,
with $\lambda$  expressed in $\mu$m.   So for the 14.3  $\mu$m filter,
the full  width of the  first Airy minimum  was 12$''$ resulting  in a
Rayleigh  criterion of 6$''$.   The corresponding  number for  the 7.7
$\mu$m filter was 3.2$''$.  All CAM data processing was performed with
the CAM Interactive Analysis (CIA) system \citep{ott97,del00}, and the
following  method  was  applied   during  data  processing:  (i)  Dark
subtraction was performed using a  dark model with correction for slow
drift of the dark current throughout the mission.  (ii) Glitch effects
due to cosmic rays were removed  following the method of Aussel et al.
(1999).  (iii)  Transient correction for  flux attenuation due  to the
lag in the detector response  was performed by the method described by
Abergel et al.  (1996).  (iv)  The raster map was flat-fielded using a
library flat-field.  (v) Pixels affected by glitch residuals and other
persistent effects  were manually suppressed.  (vi)  The raster mosaic
images  were  deconvolved  with  a multi-resolution  transform  method
\citep{sta98}.  Deconvolution or image restoration is a difficult step
in  image  processing especially  in  dealing  with  images with  many
instrumental  and noise  defects  such  as those  found  in CAM  data.
Consider a  signal $\mathit{I}$, the  result of the observation  of an
object $\mathit{O(x,y)}$ through an optical system. This system may be
characterised  by the point  spread function  (PSF) $\mathit{P(x,y)}$.
The observed  signal results from the degradation,  or convolution, of
$\mathit{P(x,y)}$  with $\mathit{O(x,y)}$ and  the inclusion  of noise
$\mathit{N(x,y)}$:

$\mathit{I(x,y) = (O * P)(x,y) + N(x,y)}$

To   correct   for   the    degradation   of   the   optical   system,
$\mathit{O(x,y)}$  must be  determined  knowing $\mathit{I(x,y)}$  and
$\mathit{P(x,y)}$.   The presence  of  the additive  noise means  that
there is an infinite number of solutions to this equation.  One of the
most  widely  used  techniques  for  determining  a  solution  is  the
multi-resolution transform  method \citep{sta98}.  It is  based on the
idea that a  given realization of a random  variable carries an amount
of information  that is inversely proportional to  its entropy.  Among
an infinite  number of solutions,  the one that maximises  its entropy
will minimise the amount of information required to make it compatible
with  the  observed  data.   The deconvolved  map  yielded  resolution
enhancements  of  a  factor of  1.2  to  1.6  with the  higher  values
associated with  the 7.7  micron map. Spatial  resolution for  the 7.7
micron map  was 2.2$''$, and 5$''$  for the 14.3  micron map.  Similar
resolution enhancements by  factors of 1.2 to 2.0  were obtained using
the  multi-resolution  transform  method  on  HST and  also  ISO  data
\citep{wu94,pan96}.  There was a loss  of flux in this process for the
weaker, point-like  sources (D, E).  On the other hand,  more extended
sources  such as  the galaxies  A and  C have  their  fluxes conserved
through the deconvolution process.  The deconvolved maps are presented
here,  with the  source  photometry performed  on the  non-deconvolved
maps.  Colour ratios  taken on the deconvolved maps  are valid because
any attenuation of source flux  occurs similarly for a given source in
each  filter.   Photometry was  performed  by  integrating the  pixels
containing source flux exceeding  the background by 3$\sigma$.  Fluxes
measured  in broadband filters  are essentially  valid either  for the
nominally assumed  $\lambda$\(^{-1}\) source spectrum, or  even if the
measured  and modelled galaxy  spectrum is  taken into  account, since
colour correction terms are negligible in this case, compared with the
measurement  precision.   The lowest  contour  level is  approximately
twice the  standard deviation of  the background noise  obtained using
pixels  from  around the  border  of  the  image.  The  ISOCAM  fluxes
presented in  Table 2  have a photometric  accuracy of about  15\%.  A
comparison   of  photometry   for   each  source   before  and   after
deconvolution is given in Table 3.   IRAS fluxes for NGC 1741 are also
given for comparison in Table 4 \citep{mos90}.  The difference between
the 14.3 $\mu$m  and the IRAS 12 $\mu$m fluxes is  probably due to the
strong 8.5 and 11.3 $\mu$m  PAH features in the IRAS band.  Comparison
of  images  from  different  filters  was  obtained  by  matching  the
resolution of  the two deconvolved images before  creating ratio maps.
The durations  of the observations in  Table 1 give the  length of the
actual  observation   including  instrumental,  but   not  spacecraft,
overheads.

\subsection{PHT-S}

 PHT-S consists of a dual  grating spectrometer with a resolving power
of 90  \citep{lau98}.  Bands SS  and SL cover  the range of 2.5  - 4.8
$\mu$m and 5.8 - 11.6  $\mu$m respectively.  The PHT-S spectrum of the
nucleus  of NGC 1741  was obtained  by operating  the $24''\times24''$
aperture of PHT-S in rectangular chopping mode using the ISOPHOT focal
plane chopper.   In this mode,  the satellite pointed to  the position
between the source  and an off-source position, and  the chopper moved
alternately between these two positions.  The source was always in the
positive beam  in the spacecraft Y-direction.  For  the PHT-S spectrum
of the lobe  of galaxy A, PHT-S was operated in  a similar manner.  In
Figure 1 the nominal PHT-S aperture (24"x24" square) is shown overlaid
on the  7.7 $\mu$m map, for  the case of  the off-nucleus measurement.
However, the beam profiles differ in detail per PHT-S detector and are
in  general narrower  than  the nominal  aperture \citep{lau01}.   The
calibration of the spectrum was performed by using a spectral response
function   derived  from  several   calibration  stars   of  different
brightness  observed  in  chopper  mode \citep{aco99}.   The  relative
spectrometric  uncertainty of the  PHT-S spectrum  is about  20\% when
comparing different  parts of  the spectrum that  are more than  a few
microns apart.  The absolute photometric uncertainty is about 30\% for
bright  calibration sources.   All PHT  data processing  was performed
using  the  ISOPHOT Interactive  Analysis  (PIA)  system, version  8.1
\citep{gab98}.  Data  reduction consisted primarily of  the removal of
instrumental effects such  as radiation events which result  in a jump
of  a minimum  of two  consecutive read-out  values, i.e.   in  a step
within the  integration ramp.  The  disturbance is usually  very short
and the  slope of the  ramp after the  glitch is similar to  the slope
before it.  Once the instrumental effects had been removed, background
subtraction  was performed  and flux  densities were  obtained.  These
fluxes were  plotted to obtain  the spectra for the  central starburst
region of NGC 1741 and the nucleus of HCG 31A.

\section{Results}

A map of  the HCG 31 group consisting of a  deconvolved 7.7 $\mu$m LW6
map overlaid  on a  red Deep Sky  Survey (DSS)  image of the  group is
presented in  Fig.  1.   Deconvolved 7.7 and  14.3 $\mu$m maps  of the
central galaxies overlaid on DSS and H$\alpha$ images are presented in
Figs.  2 and 3.

The bulk  of the 7.7  $\mu$m and 14.3  $\mu$m emission comes  from the
interacting galaxies  A and  C, which are  together designated  as NGC
1741.  Peak emission,  which is  denoted as N  on each  map, coincides
with the central  region of galaxy C.  A  secondary peak is associated
with the nucleus  of galaxy A, and forms an  extension to the emission
from galaxy C.   The emission at 14.3 $\mu$m  is primarily confined to
NGC 1741, though galaxy E and  a star forming knot within galaxy B are
also  weakly detected.   Since emission  at 14.3  $\mu$m is  a primary
tracer   of  warm   dust  consisting   of  very   small   dust  grains
stochastically heated in regions  of low excitation and/or heated dust
close to luminous  stars, this indicates that most of  the dust in the
system is within the galaxy NGC 1741.  On the 7.7 $\mu$m map, emission
is much  more widespread,  with galaxies B  and E well  detected along
with  two sources  forming an  extension  of emission  from galaxy  A.
Since  the 7.7  $\mu$m PAH  feature  is indicative  of star  formation
\citep{cha99}, this  map provides a  clear representation of  the star
formation within  the system,  a fact outlined  by the  good agreement
between  the 7.7  $\mu$m  contours and  the  underlying H$\alpha$  map
\citep{igl97}.  However, a surprising finding was the non-detection of
galaxy F on the 7.7 $\mu$m map, given its easy detection in H$\alpha$.
Two other  sources were also  detected on the  7.7 $\mu$m map  and are
denoted by HGC  31:IR1 and HGC 31:IR2.  These sources  do not have any
known optical  or H$\alpha$ counterparts, and a  careful inspection of
the data cube before coadding  of all map pointings suggests that they
are  the residue  of  cosmic  ray glitches  that  were not  completely
eliminated, and  we are  not treating them  as confirmed  sources.  As
regards the remaining  members of the group, G and  Q were outside the
ISOCAM map and  thus not observed (Fig.  1).   The background galaxy D
was detected at 7.7 $\mu$m, but not in H$\alpha$.  The Tycho catalogue
star TYC  04758-466-1, with  m$_{V}$ = 11.3  and spectral type  F6 was
detected  at 7.7$\mu$m  and 14.3  $\mu$m, with  a flux  ratio  that is
consistent with  the Rayleigh-Jeans tail of a  blackbody, which serves
as  a valuable cross-check  on the  ISOCAM photometric  calibration at
these  faint fluxes.   The  ISOCAM  fluxes for  each  source from  the
non-deconvolved maps are given in Table 2.

A  PHT-S spectrum  of  the central  starburst  region of  NGC 1741  is
presented in  Fig.  4.  Detections were made  of unidentified infrared
bands  (UIBs)  at 6.2,  7.7,  8.6 and  11.3  $\mu$m  that are  usually
attributed to  PAHs.  The 8.6 $\mu$m  feature has a  peak flux roughly
equal  to  the 7.7  $\mu$m  feature,  which  is unusual  in  starburst
spectra.   Such  features may  be  indicative  of  newly created  PAHs
\citep{ver96}.  The feature at 10.65 $\mu$m may be a blend of features
including [S  IV].  A  PHT-S spectrum  of the nucleus  of galaxy  A is
presented in  Fig.  5. PAHs  at 6.2, 7.7,  8.6, and 11.3 $\mu$m  and a
feature at 9.7$\mu$m, possibly PAH,  were detected but at a level much
weaker than  in the starburst  region.  For both spectra,  line fluxes
for each feature were  obtained by interpolating the continuum between
the ends  of the  wavelength range.  The  line fluxes for  the central
starburst  region are  given  in Table  5,  while the  fluxes for  the
nucleus of galaxy A are given in Table 6.

The spectral energy distribution of NGC 1741 using ISOCAM, PHT-SL from
the central  starburst and IRAS fluxes  is presented in  Fig.  6.  The
dust model, denoted by the solid curve \citep{kru94,sieb98} was fitted
to  the data  and contains  three separate  dust populations:  the PAH
bands \citep{bou98}, a warm dust component  at 110 K and a cooler dust
component at 40 K, and each component is indicated in Fig.  6.

\section{Discussion}

\subsection{Morphology and star formation in HCG 31}

The interaction between galaxies A and C has been the primary cause of
the  severely disturbed morphology  and high  rates of  star formation
within the group.  The  most obvious morphological features created by
the interaction are the tidal tails; the northern tidal tail extending
from  A and  containing two  ISO sources,  and a  southern  tidal tail
containing dwarf galaxies E and F.  The stripping of H I from galaxies
A and  C allowed the  dwarf galaxies to  condense and form  within the
southern tidal tail.  Galaxies B and  G have also been affected by the
central  interaction of A  and C.   Galaxy B  appears quite  warped in
optical  and continuum maps  \citep{rub90,igl97}, with  star formation
located in three H$\alpha$ knots.  The galaxy is also linked to galaxy
C  by a  bridge detectable  in H$\alpha$.   G exhibits  star formation
weighted  towards the  edge of  the galaxy  facing the  southern tail,
indicating that star formation within the galaxy has been triggered by
interaction  with  the  tail.   The  peculiar  system  morphology  and
resultant star  formation led to investigations of  the star formation
history of the  group \citep{igl97} with three distinct  modes of star
formation suggested.  The strongest  mode of star formation, involving
galaxies A and C, was a product of the collision between the two, with
star formation being concentrated in two strong bursts coinciding with
the nuclei  of the galaxies.   The secondary mode, directly  linked to
this interaction,  involves the formation  of dwarf galaxies from  H I
stripped  from both  galaxies.  Galaxies  E and  F seem  to  have been
formed in  this manner.   Finally, a third  mode shows  star formation
displaced along  the edges of  the affected galaxies,  concentrated in
knots.  This mode, responsible for star formation in galaxies B and G,
may be  due to a  stripping mechanism as  the galaxies pass  through a
denser medium  such as H I stripped  from galaxies A and  C.  Given an
interaction timescale of about 400 Myr and an age of the central burst
of about  4 to 5  Myr \citep{joh99}, it  is clear that these  modes of
star  formation are  very  recent, ongoing,  events  in the  dynamical
history of HCG 31.  However, the occurance of previous periods of high
star formation may be noted  by the presence of stellar super-clusters
in the central burst with ages over 10 Myr \citep{joh99,joh00}.

NGC  1741   was  previously   surveyed  in  X-ray   \citep{ste98},  UV
\citep{con96},     optical/H$\alpha$     \citep{rub90,igl97},    radio
\citep{wil91} and CO \citep{yun97}.  It has been estimated that $\sim$
700 Wolf Rayet and $\sim$ 10$^{4}$ O-type stars are located within the
central burst  \citep{con96}.  With some  knowledge of the  content of
the burst,  it is instructive to  determine the current  state of star
formation in NGC 1741 using  a diagnostic such as the 14.3/6.75 $\mu$m
flux  ratio for  the  system.  The  14.3/6.75  $\mu$m ratio  generally
decreases    as    interactions    develop    and    starbursts    age
\citep{vig99,ces99,cha99,hel99}.   To determine  the  14.3/6.75 $\mu$m
ratio  for NGC  1741, the  6.75 $\mu$m  (ISOCAM LW2  filter)  flux was
synthesized from  the PHT-S spectrum  in the wavelength range  of that
ISOCAM filter  (5 to 8.5 $\mu$m).   In order to check  the accuracy of
the  synthesised  flux,  6.75   $\mu$m  and  7.7  $\mu$m  fluxes  were
synthesised  for  several sources,  including  Mkn  297  and NGC  7714
\citep{ohal00}, from their PHT-S  spectra, and the derived values were
found  to agree  with  the  directly measured  ISOCAM  fluxes for  the
sources  within  the photometric  uncertainties.   For  NGC 1741,  the
14.3/6.75 $\mu$m  ratio was  found to be  1.6$\pm$0.4.  This  value is
quite  low. A  ratio  above 3  is  indicative of  high star  formation
leading to  the heating of  dust in the  nuclear region by  hot, young
ionising stars \citep{vig99,lau00}.

Since the  14.3 $\mu$m  flux is dominated  primarily by  emission from
very  small  dust  grains and  the  6.75  and  7.7 $\mu$m  fluxes  are
dominated  by PAHs, the  14.3/7.7 $\mu$m  ratio provides  a diagnostic
similar to the 14.3/6.75 $\mu$m ratio, allowing a determination of the
current  state of  the starburst.   To determine  the  14.3/7.7 $\mu$m
ratio, fluxes were obtained in strips  on each of the raster maps with
each strip  bisecting NGC 1741  through the central  starburst region.
Each  strip  was  equally  separated  by  a  position  angle  (PA)  of
45$^{\circ}$, measured from north, where PA = 0$^{\circ}$.  The strips
were  divided   to  obtain  the  14.3/7.7  $\mu$m   ratio.   One  such
ratio-strip is presented in Fig.  7, taken from northwest to southeast
through the  galaxy.  The slope  of the 14.3/7.7 $\mu$m  ratio changes
over the length  of the strip and  attains a peak value of  1.9 at the
center  of the starburst.   To the  northeast of  the burst  the ratio
drops to 1.5, while to the  southwest the slope is steeper, reaching a
minimum  of  1.3.   The higher  value  of  the  ratio in  the  central
starburst region  may be due to  destruction of PAHs  by UV ionisation
from O  and B type  stars, with an  added component due to  heating of
very small dust  grains.  In comparison with other  galaxies with very
active starbursts where  the 14.3/6.75 $\mu$m ratio can  exceed 3, the
peak 14.3/7.7  $\mu$m ratio  value of  1.9 for NGC  1741 is  quite low
\citep{cha99}.  The  low  ratio  values  and  the  age  of  the  burst
\citep{joh00} suggest that, while the small central region of NGC 1741
is still in the throes of  a very recent starburst, the central region
may  be moving  into  a  post-starburst stage  with  cessation of  the
formation  of  the  most  massive  stars \citep{cha99}.   This  is  in
contrast to  previous higher  resolution surveys which  suggest strong
ongoing    star    formation,    especially   in    compact    regions
\citep{rub90,con96,igl97,joh00}. However,  the lower ratio  values for
NGC 1741 may  be due to ISO being sensitive to  a more extended region
with less activity.   The aging of the burst may  be due to supernovae
and  stellar winds \citep{tan88}  disrupting the  interstellar medium,
preventing further star formation as soon as the first generation of O
stars have formed and evolved \citep{gen00}.

\subsection{Does HCG 31 harbour an AGN?}

Several surveys of  HCGs have led to a debate about  the nature of the
activity  within  compact  groups.   Sulentic and  Rabaca  (1993)  and
Venugopal (1995) contest the claim  by Hickson et al.  (1989) that far
infrared emission  is enhanced in  compact groups, arguing  that while
tidal interactions and mergers did occur in the past in compact groups
\citep{rub91,men95}, most  compact group galaxies seem  to exhibit the
behaviour of  normal galaxies.   However, a significant  proportion of
the brightest HCGs  do display some sort of  activity, be it starburst
or AGN \citep{coz98a}.

CO surveys  of HCGs \citep{yun97,ver98}  have shown that  CO deficient
groups are  also the  most dynamically evolved  and show a  history of
strong star formation.   An extreme example is HCG  16, which includes
one  Seyfert 2  galaxy, two  luminous LINER  galaxies and  3 starburst
galaxies \citep{rib96}.  Those groups with AGN components tend to have
their AGN  concentrated within the  group core.  This is  likely since
AGN tend to be found within luminous and early type galaxies.

Several  diagnostic tools  are available  to probe  the nature  of the
activity within the  central region.  The ratio of  the integrated PAH
luminosity  and  the  40  to  120 $\mu$m  IR  luminosity  \citep{lu99}
provides  a tool to  discriminate between  starbursts, AGN  and normal
galaxies because the lower the ratio, the more active the galaxy.  For
NGC 1741,  the ratio is 0.08  and is consistent with  the values found
for other  starbursts \citep{vig99}.  Similarly, the ratio  of the 7.7
$\mu$m PAH flux to the  continuum level at this wavelength can provide
a   measure   of   the   level   of  activity   within   the   nucleus
\citep{gen98,lau00,cla00}.   The ratio  for  NGC 1741  is  3.5 and  is
indicative of an active  starburst.  The results from these diagnostic
tools  indicate that  NGC 1741  is  home to  a compact  burst of  star
formation and that an AGN is not required within the central region of
the merging galaxies HCG 31 A and C.

\subsection{How will HCG 31 evolve?}

Coziol et  al.  (1998b) discuss  a possible galaxy  evolution scenario
for the HCGs.   The first phase involves triggering  of starbursts due
to mergers, as in HCG 31.  When the galaxy's mass is high enough and a
large enough gas reservoir is available, the gas falls inwards towards
the centre  to form and nourish  an AGN.  As the  starburst fades, the
AGN can  remain as a Seyfert or  LINER depending on the  amount of gas
available  and the  level of  accretion.  When  the gas  reservoir has
finally dissipated, a low luminosity AGN or a normal galaxy may be the
final result.  Most HCGs are  dominated by such low luminosity AGNs or
non-active galaxies in their cores indicating that the activity in the
system has passed and they are now in a quieter phase.

As  regards the  eventual  fate of  HCG  31, models  by Barnes  (1989)
suggest that  after continuing  mergers, the final  result will  be an
elliptical galaxy.   After star formation activity has  ceased and the
remaining stars have evolved,  the luminosity will fade \citep{rub90}.
The HCG  31 group  sits within  a large common  H I  envelope totaling
$\sim$  10$^{10}$ M$_{\odot}$  of gas,  which  may end  up within  the
center of the  object where it can fuel a starburst  or an AGN.  Given
such a large gas reservoir, and  the lack of an AGN within the central
burst of HCG 31,  it would seem that HCG 31 is  in the early stages of
such a scenario.

\section{Conclusions}

  The 14.3/6.75 $\mu$m and 14.3/7.7 $\mu$m flux ratios (where the 6.75
$\mu$m flux was synthesized  from PHT-S measurements) suggest that the
central burst within NGC 1741 may be moving towards the post-starburst
phase.  The  ratio of the integrated  PAH luminosity to the  40 to 120
$\mu$m infrared  luminosity and  the far-infrared colours  reveal that
despite the high surface brightness  of the nucleus, the properties of
HCG 31 can be explained in terms of a starburst and do not require the
presence an  AGN. The  star TYC 04758-466-1,  with m$_{V}$ =  11.3 and
spectral type F6 was detected at 7.7$\mu$m and 14.3 $\mu$m.

\section{Acknowledgements}

We  thank J.   Iglesias-Paramo  for  kindly allowing  the  use of  the
H$\alpha$ map of HCG 31.  The ISOCAM data presented in this paper were
analysed  using  CIA, a  joint  development  by  the ESA  Astrophysics
Division and the  ISOCAM Consortium.  The ISOCAM Consortium  is led by
the ISOCAM PI, C.  Cesarsky.  The ISOPHOT data presented in this paper
were  reduced using  PIA,  which is  a  joint development  by the  ESA
Astrophysics   Division   and   the   ISOPHOT  consortium   with   the
collaboration   of  the  Infrared   Processing  and   Analysis  Center
(IPAC). Contributing  ISOPHOT Consortium  members are DIAS,  RAL, AIP,
MPIK, and MPIA.

\begin{figure} \resizebox{\columnwidth}{!}  {\includegraphics*{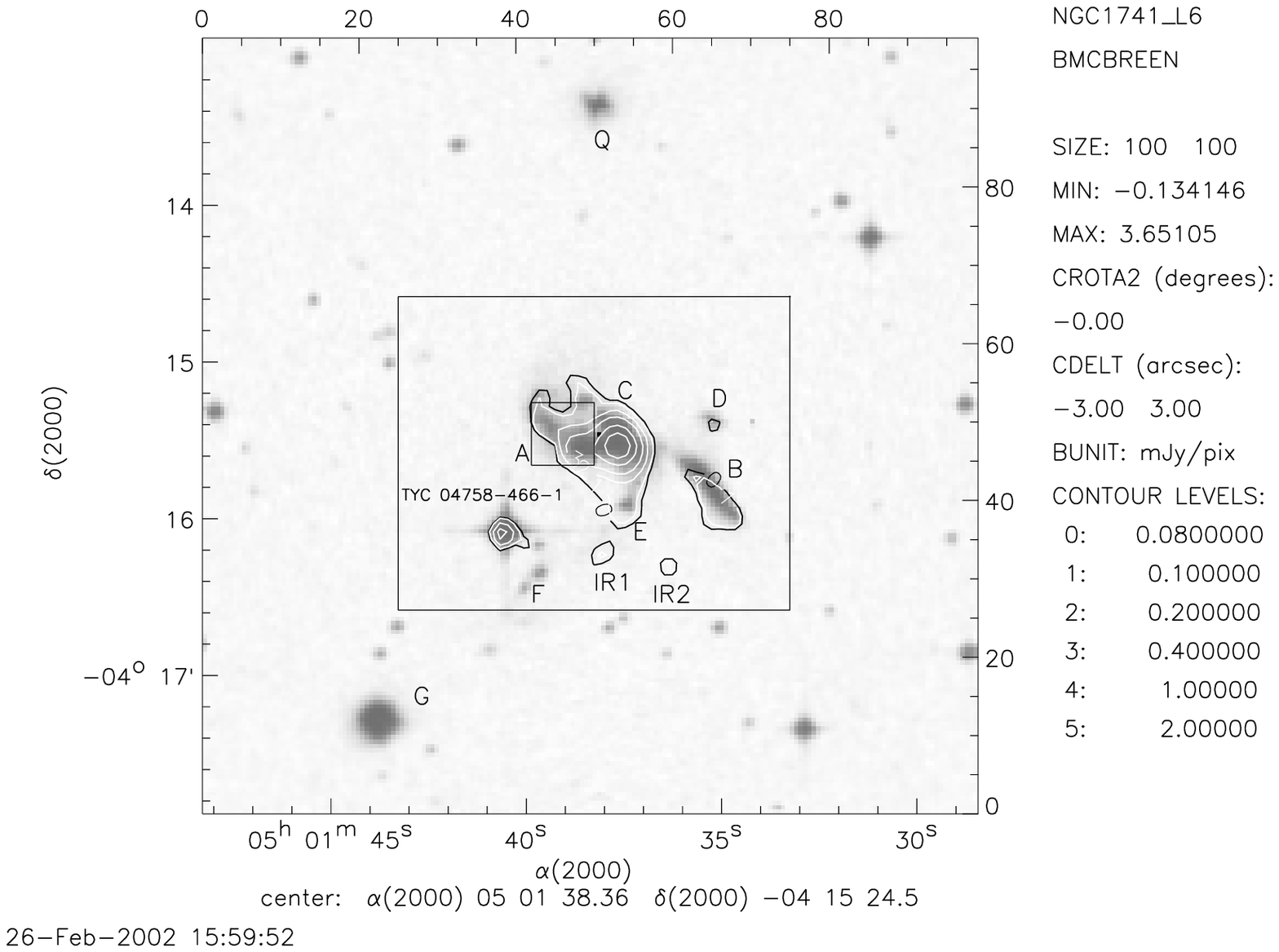}} 

\caption{Deconvolved 7.7$\mu$m map of the Hickson 31 group overlaid on
a red DSS image.  Each member of the  group (A to C, E to G, and Q) is
identified, along with  background galaxy D.  North is  to the top and
east to  the left.   The large box  delineates the region  mapped with
ISOCAM, while the small box indicates  the position and size of one of
the ISOPHOT beams.  The weak sources  HCG 31:IR1 and HCG 31:IR2 do not
have  any  known optical  or  H$\alpha$  counterparts,  and a  careful
inspection  of the  data cube  before  coadding of  all map  pointings
suggests that  they are the residue  of cosmic ray  glitches that were
not completely  eliminated and we  are not treating them  as confirmed
sources.  The top and right-hand  scales refer to a local pixel number
scale.}
\end{figure}

\begin{figure} \resizebox{\columnwidth}{!}  {\includegraphics*{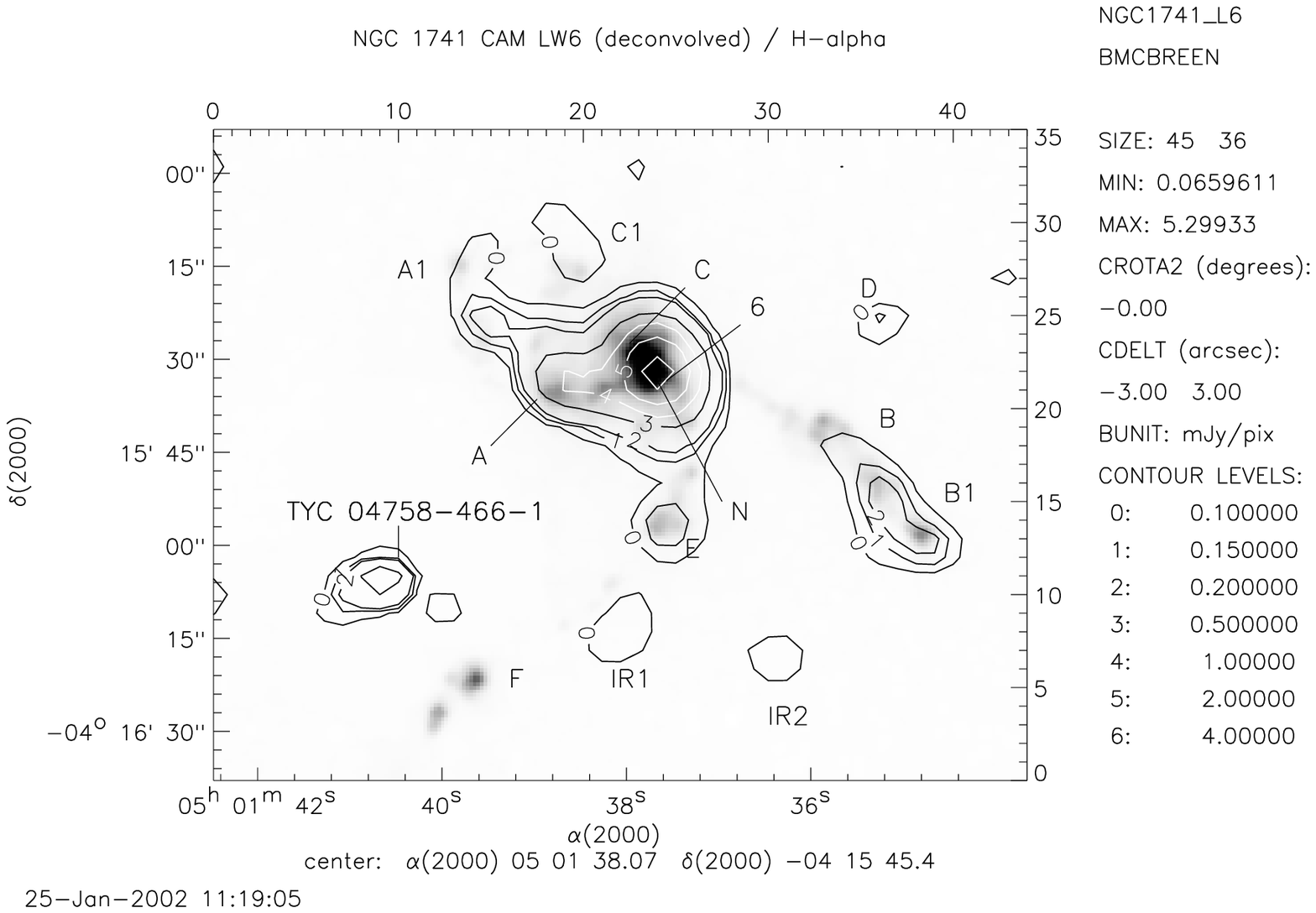}} 

\caption{Deconvolved  7.7$\mu$m  map  of  HCG 31,  superimposed  on  a
H$\alpha$ map of the group.  Contour levels (mJy/arcsec\(^{-2}\)) are:
0 = 0.1; 1 = 0.15;  2 = 0.2; 3 = 0.5; 4 = 1.0; 5 =  2.0; 6 = 4.0. N is
the starburst  region. Sources  A1 and C1  are extensions  of emission
from  galaxy A  and  lie in  the  northern tidal  tail.  Source B1  is
associated with the southernmost H-alpha knot within galaxy B. The map
notation is the same as in Fig.  1.}
\end{figure}

\begin{figure} \resizebox{\columnwidth}{!}  {\includegraphics*{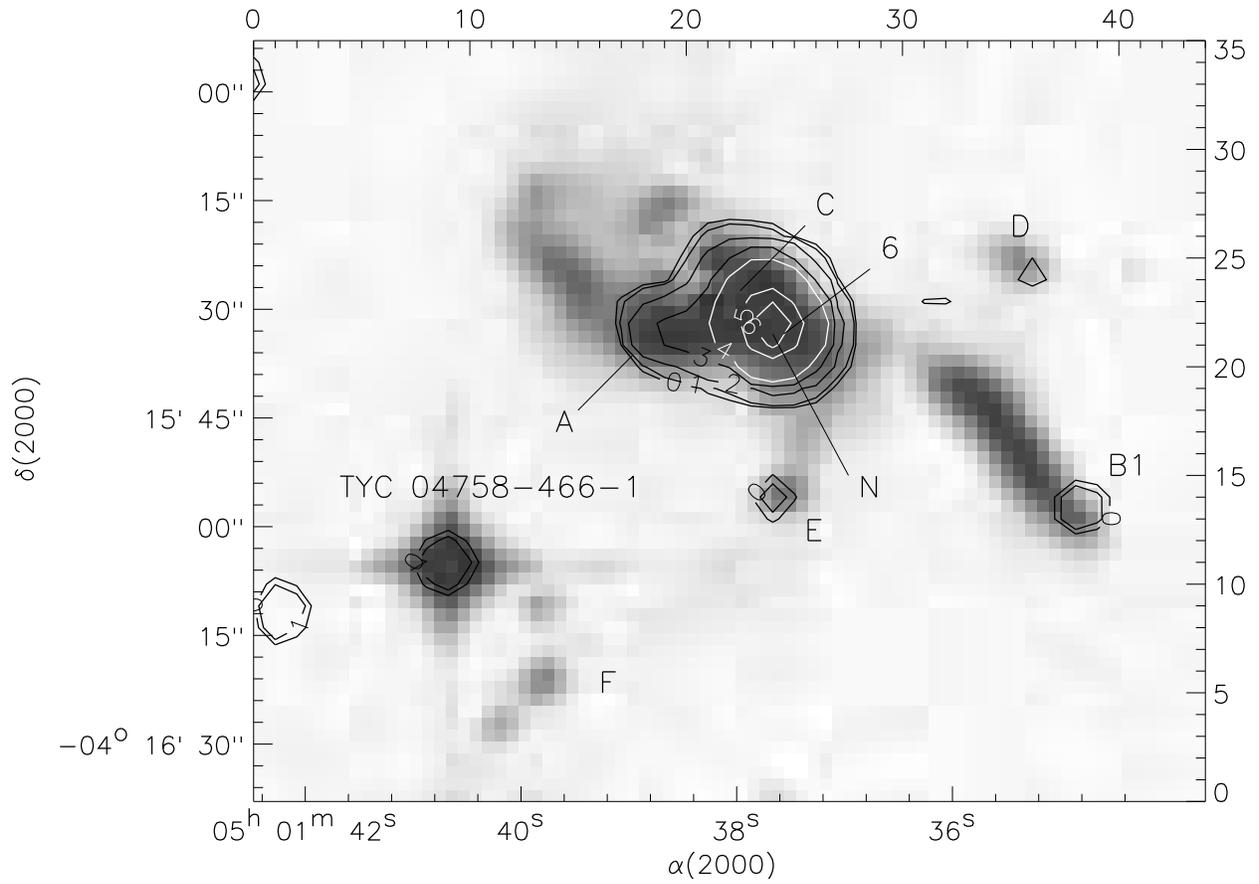}} 

\caption{Deconvolved 14.3 $\mu$m map of  HCG 31, superimposed on a red
DSS image of the  system.  Contour levels (mJy/arcsec\(^{-2}\)) are: 0
= 0.1; 1  = 1.0; 2 = 2.0;  3 = 4.0; 4 =  6.0, 5 = 8.0, 6  = 10.0.  The
galaxy identifications are otherwise the same as in Figs.  1 and 2.}
\end{figure}

\begin{figure} \resizebox{\columnwidth}{!}  {\includegraphics*{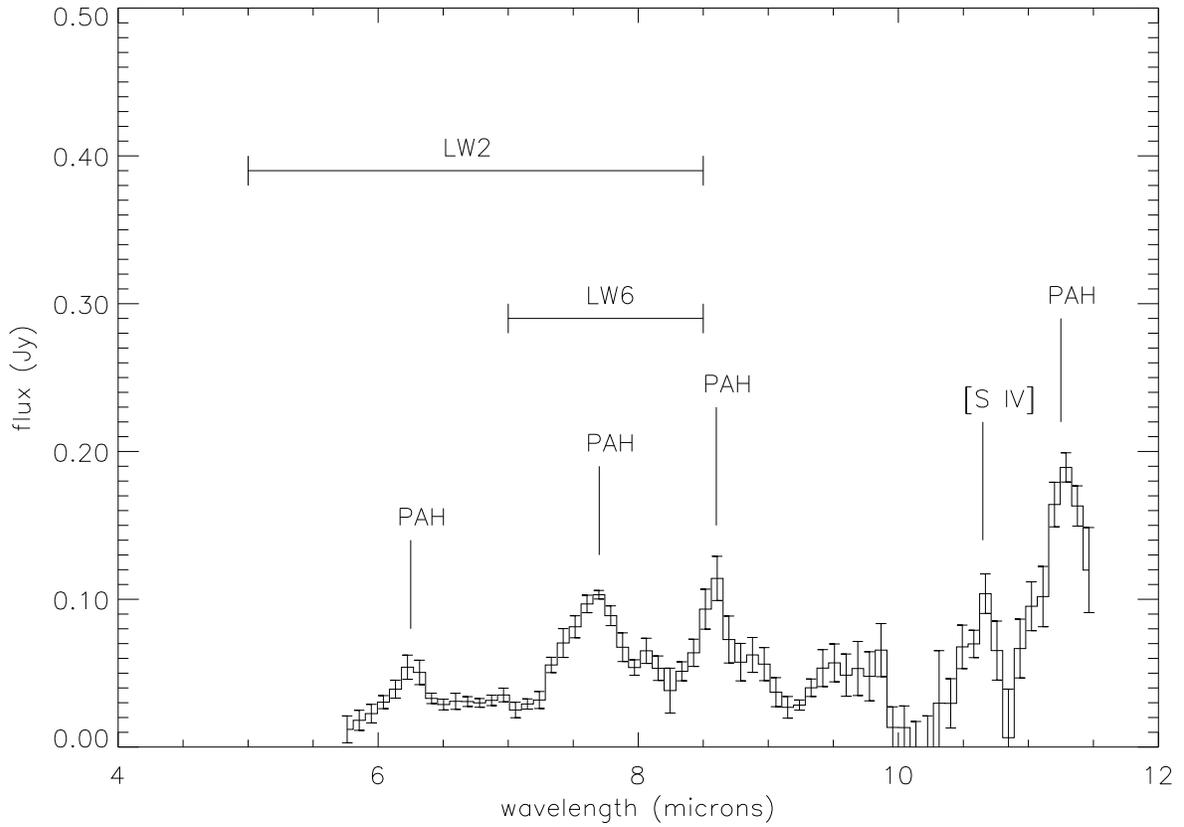}}
\caption{PHT-S  spectrum of  the central  starburst region of NGC 1741.
The PAH  and [S  IV] features are  indicated.  The bandpasses  for the
6.75  $\mu$m   and  7.7  $\mu$m   ISOCAM  filters  are   also  shown.}
\end{figure}

\begin{figure} \resizebox{\columnwidth}{!}
{\includegraphics*{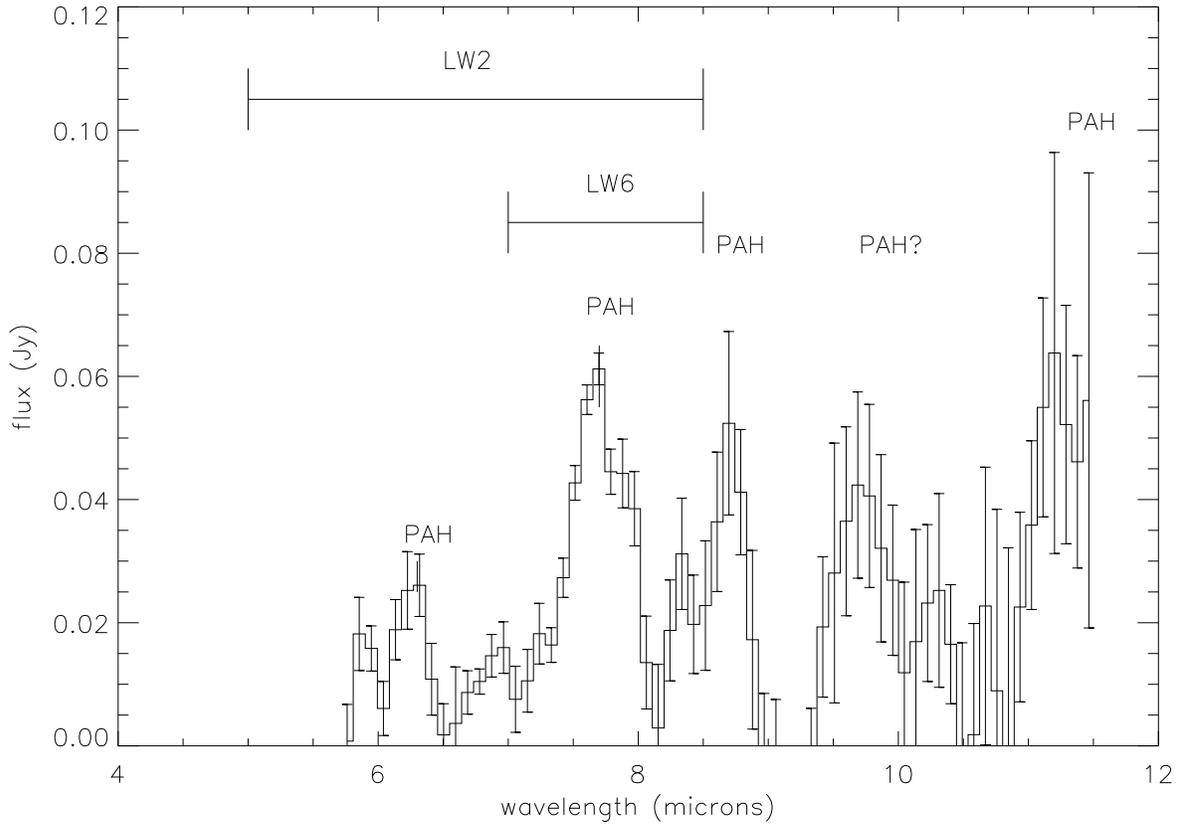}} 

\caption{PHT-S spectrum  of the  nucleus of galaxy  HCG 31A.   The PAH
features are  indicated.  The bandpasses  for the 6.75 $\mu$m  and 7.7
$\mu$m ISOCAM filters are also given.}

\end{figure}

\begin{figure} \resizebox{\columnwidth}{!}  {\includegraphics*{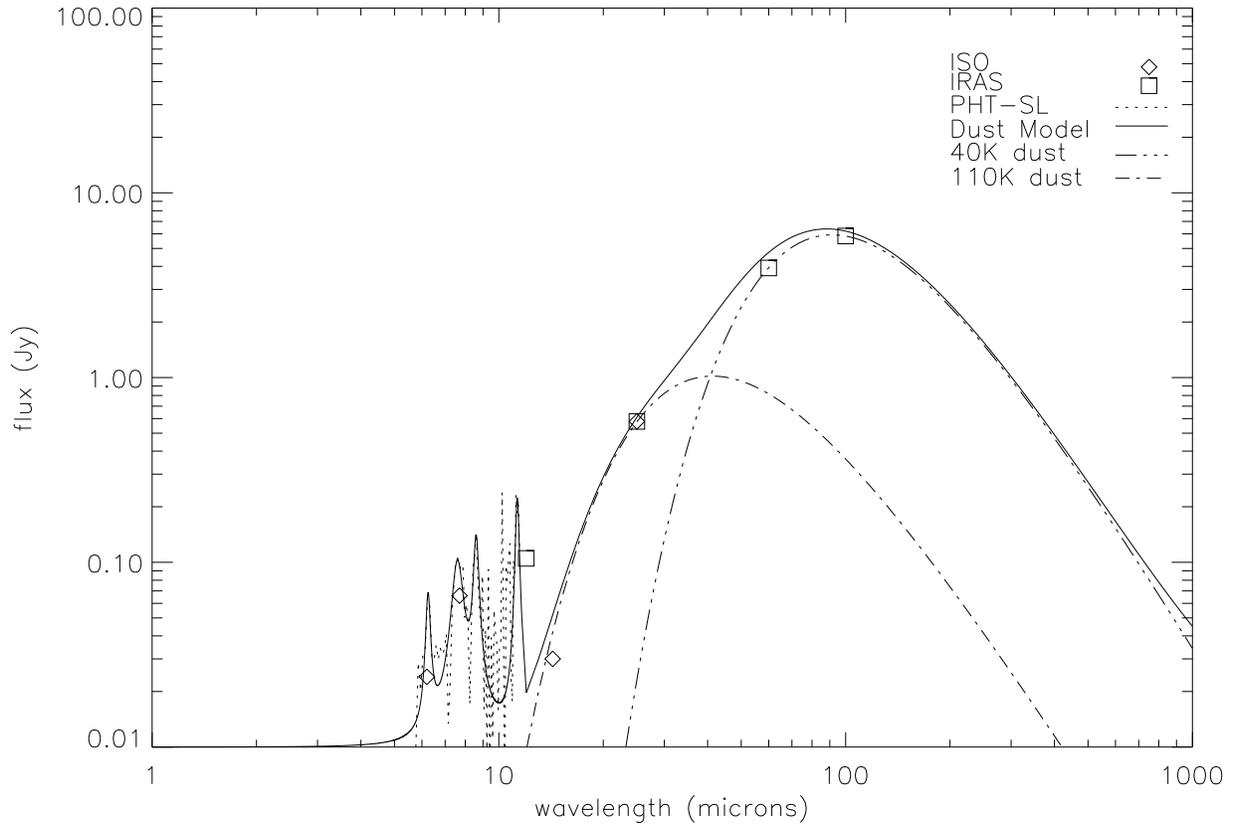}}

\caption{Spectral energy  distribution of NGC 1741 using  ISO and IRAS
flux data, including the key for the different symbols. The PHT-S data
is taken  from the central starburst.   The models fitted  to the data
are described in the text.}

\end{figure}

\begin{figure} \resizebox{\columnwidth}{!}  {\includegraphics*{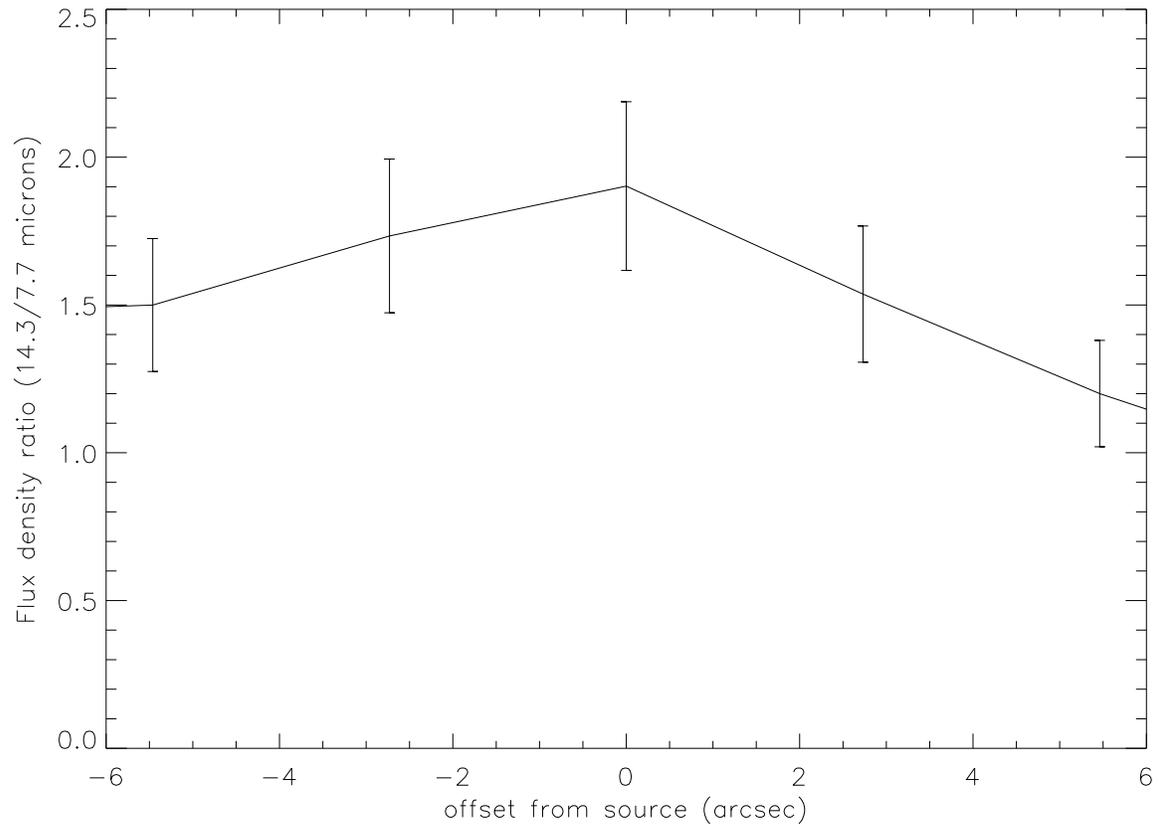}} 

\caption{14.3/7.7  $\mu$m ratio of  ISOCAM emission  profiles obtained
from the NE to SW scan across the central starburst region of NGC 1741
at PA = 45$^{\circ}$.}

\end{figure}

\clearpage
\begin{deluxetable}{cccccccccc}\rotate
\tabletypesize{\scriptsize} 

\tablecaption{ Log of the ISO observations of HCG 31.  The ten columns
list the  observation name, the TDT  number, the AOT,  the filter, the
wavelength  range ($\Delta\lambda$), the  reference wavelength  of the
filter, the date and duration  of the observation, and the position of
the observation in Right  Ascension and Declination respectively.  The
+ symbol represents the central  starburst region of NGC 1741, while *
represents the nucleus of HCG 31A.  \label{tbl-1}}

\tablewidth{0pt} \tablehead{\colhead{Observation}
&  \colhead{TDT number}  & \colhead{AOT  number} &  \colhead{Filter} &
\colhead{$\Delta\lambda$}    &    \colhead{$\lambda_\mathrm{ref}$}   &
\colhead{Date}  &  \colhead{Duration}  & \colhead{Right  Ascension}  &
\colhead{Declination}\\ & &  & & \colhead{($\mu$m)} & \colhead{($\mu$m)}
&  & \colhead{(seconds)} &\colhead{(RA)}  &\colhead{(DEC)}} \startdata
NGC1741 LW3 &85801324 &CAM01 &LW3 &12.0--18.0 &14.3 &22 Mar 1998 &2076
&05\(^{h}\)   01\(^{m}\)  37.7\(^{s}\)  &-04$^{\circ}$   15$'$  32.4$''$\\
NGC1741 LW6  &85800809 &CAM01 &LW6  &7.0--8.5 &7.7 &22 Mar  1998 &2176
&05\(^{h}\)   01\(^{m}\)  37.7\(^{s}\)  &-04$^{\circ}$   15$'$  32.4$''$\\
NGC1741 +  &68801908 &PHOT40  &SS/SL &2.5--11.6 &  &04 Oct  1997 &1132
&05\(^{h}\)   01\(^{m}\)  37.7\(^{s}\)  &-04$^{\circ}$   15$'$  31.7$''$\\
NGC1741 *  &85800921 &PHOT40  &SS/SL &2.5--11.6 &  &22 Mar  1998 &1132
&05\(^{h}\)   01\(^{m}\)  38.9\(^{s}\)  &-04$^{\circ}$   15$'$  30.2$''$\\
\enddata
\end{deluxetable}

\clearpage
\begin{deluxetable}{crr}
\tabletypesize{\scriptsize} 

\tablecaption{ISOCAM fluxes for NGC  1741.  The three columns list the
source, the flux measured at 7.7 $\mu$m, and the flux measured at 14.3
$\mu$m.  The wavelength range for the 7.7 $\mu$m filter was 7.0 to 8.5
$\mu$m, while the range of the 14.3 $\mu$m filter was 12 to 18 $\mu$m.
The fluxes  are taken from  the non-deconvolved maps for  all objects.
The beam  size for  the ISOCAM observations  was 3$''$ pixel  field of
view (PFOV).  \label{tbl-2}}

\tablewidth{0pt}    
\tablehead{\colhead{Source}    &
\colhead{7.7 $\mu$m flux}    &    
\colhead{14.3 $\mu$m flux}\\
   &   
\colhead{(mJy)}   &
\colhead{(mJy)}}

\startdata
 
HCG 31A  & 23.20 $\pm$ 3.50 & 12.30  $\pm$ 1.90 \\ 
HCG 31B  & 5.01  $\pm$ 0.70 & $\leq$ 0.17 $\pm$ 0.00 \\
HCG 31C  & 70.80 $\pm$ 10.60 & 30.40  $\pm$ 4.60 \\ 
HCG 31D  & 0.44 $\pm$ 0.06 & $\leq$ 0.17 $\pm$ 0.00 \\
HCG 31E  & 7.86 $\pm$ 0.23 & 2.20 $\pm$ 0.42\\
TYC 04758-466-1 & 6.18 $\pm$ 1.36 & 1.53 $\pm$ 0.32\\
\enddata
\end{deluxetable}

\clearpage
\begin{deluxetable}{ccc}
\tabletypesize{\scriptsize}

\tablecaption{Comparison  of photometry  from each  source on  the 7.7
$\mu$m map before and after deconvolution \label{tbl-3}}

\tablewidth{0pt}    \tablehead{\colhead{Object}    &
\colhead{Non-deconvolved flux}    &    \colhead{Decovolved flux}\\   &   \colhead{(mJy)}   &   \colhead{(mJy)}}
\startdata
HCG 31A & 23.19 &23.17 \\ 
HCG 31B & 5.01 &4.96 \\ 
HCG 31C & 70.81 &70.77 \\ 
HCG 31D & 0.44 &0.31 \\ 
HCG 31E & 7.86 &5.22 \\ 
TYC 04758-466-1 & 6.18 &5.03 \\ 
\enddata
\end{deluxetable}

\clearpage
\begin{deluxetable}{crrr}
\tabletypesize{\scriptsize}

\tablecaption{IRAS  fluxes for NGC  1741.  The  four columns  list the
filter used,  the wavelength range,  the reference wavelength  for the
filter and the flux measured.  \label{tbl-4}}

\tablewidth{0pt}    \tablehead{\colhead{Filter}    &
\colhead{$\Delta\lambda$}    &    \colhead{$\lambda_\mathrm{ref}$}   &
\colhead{Flux}\\   &   \colhead{($\mu$m)}   &   \colhead{($\mu$m)}   &
\colhead{(mJy)}}
\startdata
IRAS 12 $\mu$m &8.5--15 &12 &110 $\pm$ 20.0 \\ 
IRAS 25 $\mu$m &19--30 &25 &580 $\pm$ 40.0 \\
IRAS 60 $\mu$m &40--80 &60 &3920 $\pm$ 310.0 \\ 
IRAS 100 $\mu$m &83--120 &100 &5840 $\pm$ 470.0 \\
\enddata
\end{deluxetable}

\clearpage

\begin{deluxetable}{crr}
\tabletypesize{\scriptsize}

\tablecaption{PHT-S  fluxes for  the central  starburst region  of NGC
1741. The three  columns give the line  identification, the wavelength
range and the integrated flux respectively. \label{tbl-5}}

\tablewidth{0pt}

\tablehead{\colhead{Line ID}   & \colhead{Wavelength range} & \colhead{Flux}\\ \colhead{($\mu$m)} & \colhead{($\mu$m)} & \colhead{(10$^{-15}$ W m$^{-2}$)}}
\startdata
PAH 6.2 & 5.8 - 6.6 & 1.25 $\pm$ 0.43 \\ 
PAH 7.7 & 7.3 - 8.4 & 2.64 $\pm$ 0.56 \\ 
PAH 8.6 & 8.4 - 9.0 & 1.40 $\pm$ 0.43 \\ 
$[$SIV$]$ 10.51 & 10.4 - 10.8 & 1.11 $\pm$ 0.17 \\ 
PAH 11.3 & 11.2 - 11.6 & 1.29 $\pm$ 0.39 \\ 
\enddata

\end{deluxetable}

\clearpage
\begin{deluxetable}{crr}
\tabletypesize{\scriptsize}

\tablecaption{PHT-S  fluxes for  the nucleus  of HCG  31A.   The three
columns  give the line  identification, the  wavelength range  and the
integrated flux respectively. \label{tbl-6}}

\tablewidth{0pt}
\tablehead{\colhead{Line ID}   & \colhead{Wavelength range} & \colhead{Flux}\\ \colhead{($\mu$m)} & \colhead{($\mu$m)} & \colhead{(10$^{-15}$ W m$^{-2}$)}}
\startdata
PAH 6.2 & 5.9 - 6.6 & 0.63 $\pm$ 0.19 \\ 
PAH 7.7 & 7.4 - 8.4 & 1.61 $\pm$ 0.44 \\ 
PAH 8.6 & 8.4 - 9.0 & 0.76 $\pm$ 0.23 \\ 
PAH 9.7?? & 9.4 - 10.0 & 0.52 $\pm$ 0.16  \\ 
PAH 11.3 & 11.2 - 11.6 & 0.47 $\pm$ 0.14 \\ 
\enddata

\end{deluxetable}

\end{document}